\documentclass[twocolumn]{aastex631}

\usepackage{rotating}
\usepackage{color}

\received{August 23, 2024}
\accepted{March 25, 2025}

\shorttitle{Unveiling chemical enrichment out to z$\sim$8}
\shortauthors{Bhattacharya et al.}

\begin{document}

\title{Unveiling galaxy chemical enrichment mechanisms out to z$\sim$8 from direct determination of O \& Ar abundances from JWST/NIRSPEC spectroscopy}

\correspondingauthor{Souradeep Bhattacharya}
\email{souradeep@iucaa.in, s.bhattacharya3@herts.ac.uk}

\author[0000-0003-4594-6943]{Souradeep Bhattacharya}
\affiliation{Inter-University Centre for Astronomy and Astrophysics, Ganeshkhind, Post Bag 4, Pune 411007, India}
\affiliation{Centre for Astrophysics Research, Department of Physics, Astronomy and Mathematics, University of Hertfordshire, Hatfield, AL10 9AB, UK}

\author[0000-0001-7214-3009]{Magda Arnaboldi}
\affiliation{European Southern Observatory, Karl-Schwarzschild-Str. 2, 85748 Garching, Germany }

\author[0000-0003-3333-0033]{Ortwin Gerhard}
\affiliation{Max-Planck-Institut für extraterrestrische Physik, Giessenbachstraße, 85748 Garching, Germany}

\author[0000-0002-4343-0487]{Chiaki Kobayashi}
\affiliation{Centre for Astrophysics Research, Department of Physics, Astronomy and Mathematics, University of Hertfordshire, Hatfield, AL10 9AB, UK}

\author[0000-0002-8768-9298]{Kanak Saha}
\affiliation{Inter-University Centre for Astronomy and Astrophysics, Ganeshkhind, Post Bag 4, Pune 411007, India}

\begin{abstract}
Galaxy chemical enrichment mechanisms have primarily been constrained by [$\alpha$/Fe] and [Fe/H] measurements of individual stars and integrated light from stellar populations. However such measurements are limited at higher redshifts (z$>1$). Recently, we proposed an analogous diagram of the oxygen-to-argon abundance ratio, log(O/Ar), vs Ar abundance, 12+log(Ar/H), as a new diagnostic window for emission nebulae. In this Letter, using line flux measurements including temperature sensitive auroral lines, we present direct determination of O and Ar abundances in nine star-forming galaxies (SFGs) from JWST/NIRSPEC spectra at z$\sim$1.3--7.7, and two more with Keck/MOSFIRE spectra at z$\sim$2.2. Utilising their positions on the log(O/Ar) vs 12+log(Ar/H) plane, we present the first inference of galaxy chemical enrichment mechanisms from an ensemble of galaxies. {Seven} SFGs at  {z$\sim$1.3--4} are consistent with the Milky Way solar neighbourhood galactic chemical enrichment models that are  driven by core-collapse and Type Ia supernovae {in a self-regulated manner}. Such enrichment mechanisms thus occur at least out to z$\sim4$. However, {four} higher-redshift SFGs (z$\sim$3.6--7.7) have {lower} log(O/Ar) values, revealing {potentially different enrichment paths} becoming important at z$>3.6$. Such log(O/Ar) values may be caused by physical mechanisms such as a rapid but intermittent star-formation and/or additional enrichment sources. This new diagnostic window for SFGs enables us to reveal the unique fingerprints of galaxy chemical enrichment out to cosmic dawn.
\end{abstract}



\section{Introduction} 
\label{sec:intro}

Since the break of cosmic dawn, the interstellar medium (ISM) of galaxies has been continually enriched by the birth and death of stars. The bulk of our understanding of galaxy chemical enrichment \citep[][]{Tinsley80,Pagel97,kob20sr, Matteucci21} stems from spectroscopic observations of individual stars in our Milky Way (MW) and studies of nearby galaxies. Deep absorption line spectra of MW stars enabled determination of their [Fe/H] and [$\alpha$/Fe], revealing the chemical composition of the ISM at the time of their birth (e.g. \citealt{Edvardsson93, Fuhrmann98, Hayden15,Imig23}). The stars showing the highest [$\alpha$/Fe] values are thought to have formed at the earliest times from ISM that had only been enriched by core-collapse (CCSNe). Once Type Ia supernovae (SNe Ia) explosions begin, more Fe is released to the ISM than previously, causing a decreasing trend in [$\alpha$/Fe] vs [Fe/H] for the subsequent generations of stars. 

To constrain early chemical enrichment mechanisms, [$\alpha$/Fe] and [Fe/H] measurements of the oldest generations of stars are required. This is also possible from  [$\alpha$/Fe] and [Fe/H] determined from integrated stellar spectra of early-type galaxies \citep[e.g.][]{Trager00,Thomas05,Kuntschner10,Greene13}. At high redshift (z$\sim2$), only a few quiescent massive galaxies have integrated stellar spectra with sufficiently deep absorption lines to enable determination of [$\alpha$/Fe] and [Fe/H] \citep{Lonoce15, Onodera15, Kriek16, Beverage24a, Beverage24b}. The vast majority of galaxies, however, are star-forming (SFGs) with their fraction increasing with increasing redshift. [Fe/H] can also be estimated from rest-frame UV continuum and combined to $\alpha$ abundance from emission lines; [$\alpha$/Fe] has been discussed for a few SFGs out to z$\sim$3.4 \citep[e.g.][]{Steidel16, Cullen19, Cullen21, Topping20, Stanton24}. 

Recently, the log(O/Ar) vs 12 + log(Ar/H) plane for emission nebulae was found to be analogous to the [$\alpha$/Fe] vs [Fe/H] plane for stars \citep{Arnaboldi22}. This was based on the analysis of emission-line spectra of planetary nebulae and HII regions surveyed in the Andromeda galaxy (M~31; \citealt{Bh+19,Bh+19b, Bh21,Bhattacharya23a}) where temperature sensitive auroral lines had been observed \citep{Bhattacharya22, Esteban20}. Like Fe, SNe Ia also preferentially produce more Ar than light $\alpha$-elements like O, whereas CCSNe produce near-constant log(O/Ar), see \citet[][]{kob20sr}. Based on this concept, \citet{Arnaboldi22} introduced the log(O/Ar) vs 12 + log(Ar/H) plane using planetary nebulae to reveal the chemical enrichment history of M~31 with high star-formation at early times ($>8$~Gyr ago) and gas-infall $\sim$2--4~Gyr ago. \citet{Kobayashi23} then showed that with well-constrained GCE models, the log(O/Ar) vs 12 + log(Ar/H) plane can be connected to the [$\alpha$/Fe]-[Fe/H] plane.

The spectra of SFGs are dominated by the emission-lines radiated by their constituent HII regions and diffuse ionised gas \citep{Sargent70}. Using abundance planes derived from their spectra such as  log(O/Ar) vs 12 + log(Ar/H) thus opens up the possibility to constrain the chemical enrichment of SFGs.

With the advent of the NIRSPEC multi-slit spectroscopy instrument on board the James Webb Space Telescope (JWST; \citealt{Jakobsen22}), direct elemental abundance determination has become possible for a number of SFGs out to z$\sim8.5$ \citep[e.g.][]{Curti23,Nakajima23,Sanders24} through the detection of the temperature sensitive auroral [OIII]$\lambda$ 4363 \AA~line. This allows the determination of abundances of a number of elements like O, Ne, S, Ar and N \citep[e.g.][]{Isobe23}. 

These elemental abundances trace the state of the cumulative chemical enrichment of its ISM by previous generations of stars. As we move to higher redshifts, the chemical abundances in galaxies increasingly map the chemical enrichment from the very early generations of stars. As the relative contribution of CCSNe and SNe Ia varies for the production of distinct elements \citep[see][]{kob20sr}, we may use the relative abundances of such elements to decipher the state of chemical enrichment of each SFG at their observed redshift.

\citet{Rogers24} and \citet{Welch24} determined log(O/Ar) values for two SFGs at {z$\sim$3 and z$\sim$1.3} respectively. \citet{Rogers24} utilised the aforementioned results \citep{kob20sr, Arnaboldi22} to interpret the state of chemical enrichment of their individual SFG with a super-solar log(O/Ar) value as being primarily enriched by CCSNe. Individual SFGs may occupy different positions in the log(O/Ar) vs 12 + log(Ar/H) plane, and thereby exhibit different states of chemical enrichment. However, by determining the positions of an ensemble of high-z SFGs in this plane, we can constrain the mechanisms that drive early galaxy chemical enrichment. 

In this work, we present the state of chemical enrichment in an ensemble of 11 SFGs at z$\sim$1.3--7.7 from their O \& Ar abundances, providing constraints on the galaxy enrichment mechanisms at these redshifts. The data and determination of O \& Ar abundances is presented in Section~\ref{sec:data1}. The positions of these galaxies in the log(O/Ar) vs 12 + log(Ar/H) plane and its implications for early galaxy chemical enrichment is  discussed in Section~\ref{sec:enrich}. We conclude in Section~\ref{sec:conc}. 


\begin{sidewaystable}
    \centering
    \setlength\tabcolsep{4pt}
    \scriptsize
    \caption{Measurements of emission line fluxes (normalised by H$\beta$ flux of 100) for the 10 galaxies in this work from their archival 1D JWST/NIRSPEC spectra.}
    \label{table:data}
    \begin{tabular}{lccccccccccccc}
    \hline
    Name & [OII] & [OII] & H$\delta$ & H$\gamma$ & [OIII] & [ArIV] & [ArIV]  & [OIII] & [OIII] & H$\alpha$  & [SII] & [SII] & [ArIII] \\
     & 3726 \AA & 3729 \AA & 4102 \AA & 4340 \AA & 4363 \AA & 4711 \AA & 4740 \AA & 4959 \AA & 5007 \AA & 6563 \AA  & 6717 \AA & 6731 \AA & 7136 \AA \\
    \hline
    ERO 10612$^a$ & 14.4 $\pm$ 3.9 & -- & 20.9 $\pm$ 3.6 & 50.8 $\pm$ 7.4 & 24.8 $\pm$ 3.7 & 14.1 $\pm$ 4.1 & 20.3 $\pm$ 5.1 & 274.7 $\pm$ 9.3 & 869.3 $\pm$ 14.9 & -- & -- & -- & -- \\
    CEERS 1536$^a$ & 63.8 $\pm$ 12.2 & -- & -- & 53.7 $\pm$ 12.1 & 32.2 $\pm$ 9.1 & -- & -- & 252.8 $\pm$ 14.3 & 795.2 $\pm$ 27.3 & 628.1 $\pm$ 37.8 & -- & -- & 31.4 $\pm$ 9.9\\
    GLASS 150029 & 17.8 $\pm$ 4.1 & 20.7 $\pm$ 4.0 & 25.5 $\pm$ 3.2 & 39.6 $\pm$ 4.6 & 18.4 $\pm$ 3.5 & -- & -- & 232.8 $\pm$ 9.1 & 658.8 $\pm$ 15.2 & 464.3 $\pm$ 11.9 & -- & -- & 16.6 $\pm$ 3.2\\
    CEERS 1665 & 51.2 $\pm$ 2.6 & 35.2 $\pm$ 2.6 & 11.8 $\pm$ 1.7 & 38.3 $\pm$ 3.4 & 6.7 $\pm$ 2.1 & -- & -- & 233.8 $\pm$ 7.9 & 733.6 $\pm$ 9.3 & 518.6 $\pm$ 8.9 & 31.0 $\pm$ 3.1 & 20.3 $\pm$ 3.9 & 14.8 $\pm$ 3.9\\
    CEERS 1651$^b$ & -- & -- & -- & 26.8 $\pm$ 5.9 & 11.4 $\pm$ 3.6 & -- & -- & 217.3 $\pm$ 8.9 & 636.6 $\pm$ 9.2 & -- & -- & -- & 28.4 $\pm$ 9.2\\
    GLASS 40066$^c$ & -- & -- & 20.1 $\pm$ 1.7 & 36.8 $\pm$ 2.8 & 7.8 $\pm$ 1.3 & -- & -- & 251.9 $\pm$ 5.5 & 768.8 $\pm$ 13.4 & 409.5 $\pm$ 6.2 & 20.2 $\pm$ 4.2 & -- & 13.3 $\pm$ 2.2\\
    JADES 19519  & 37.9 $\pm$ 6.1 & 53.3 $\pm$ 5.9 & -- & 26.2 $\pm$ 4.1 & 20.1 $\pm$ 4.0 & -- & -- & 227.9 $\pm$ 10.5 & 678.2 $\pm$ 9.2 & 511.2 $\pm$ 45.1 & 39.6 $\pm$ 8.3 & -- & 16.7 $\pm$ 3.9\\
    CEERS 11088  & 52.0 $\pm$ 3.1 & 47.0 $\pm$ 3.4 & 14.6 $\pm$ 3.9 & 38.0 $\pm$ 3.2 & 8.8 $\pm$ 2.7 & -- & -- & 193.6 $\pm$ 3.9 & 623.3 $\pm$ 7.6 & 696.6 $\pm$ 22.7 & 58.3 $\pm$ 7.1 & 25.7 $\pm$ 7.0 & 22.9 $\pm$ 5.0\\
    Q2343-D40$^c$  & -- & -- & -- & 39.7 $\pm$ 2.9 & 4.0 $\pm$ 1.6 & -- & -- & 225.4 $\pm$ 2.3 & 707.5 $\pm$ 6.9 & 555.7 $\pm$ 37.8 & 32.1 $\pm$ 2.9 & 20.4 $\pm$ 2.9 & 15.8 $\pm$ 2.1\\
    CEERS 3788  & 41.7 $\pm$ 1.9 & 45.9 $\pm$ 1.9 & 16.0 $\pm$ 2.8 & 32.2 $\pm$ 2.5 & 8.8 $\pm$ 1.6 & -- & -- & 237.0 $\pm$ 5.4 & 701.0 $\pm$ 8.2 & 608.7 $\pm$ 9.5 & 37.2 $\pm$ 5.6 & 23.3 $\pm$ 5.8 & 19.8 $\pm$ 3.5\\
    \hline
    \end{tabular}
    \tablecomments{$^a$For these galaxies, the [OII]$\lambda\lambda$ 3726, 3729 \AA~ doublet appears blended and its total flux noted in the [OII]$\lambda$ 3726 \AA~ column.\\
    $^b$For this galaxy, the [OII]$\lambda\lambda$ 3726, 3729 \AA~ doublet, H$\alpha$ and [SII]$\lambda\lambda$ 6717, 6731 \AA~ doublet lines are in chip-gaps and are hence unobserved.\\
    $^c$For these galaxies, the [OII]$\lambda\lambda$ 3726, 3729 \AA~ doublet lines are in chip-gaps and are hence unobserved.}
\end{sidewaystable}

\begin{figure*}[t]
\centering
\includegraphics[width=0.99\textwidth]{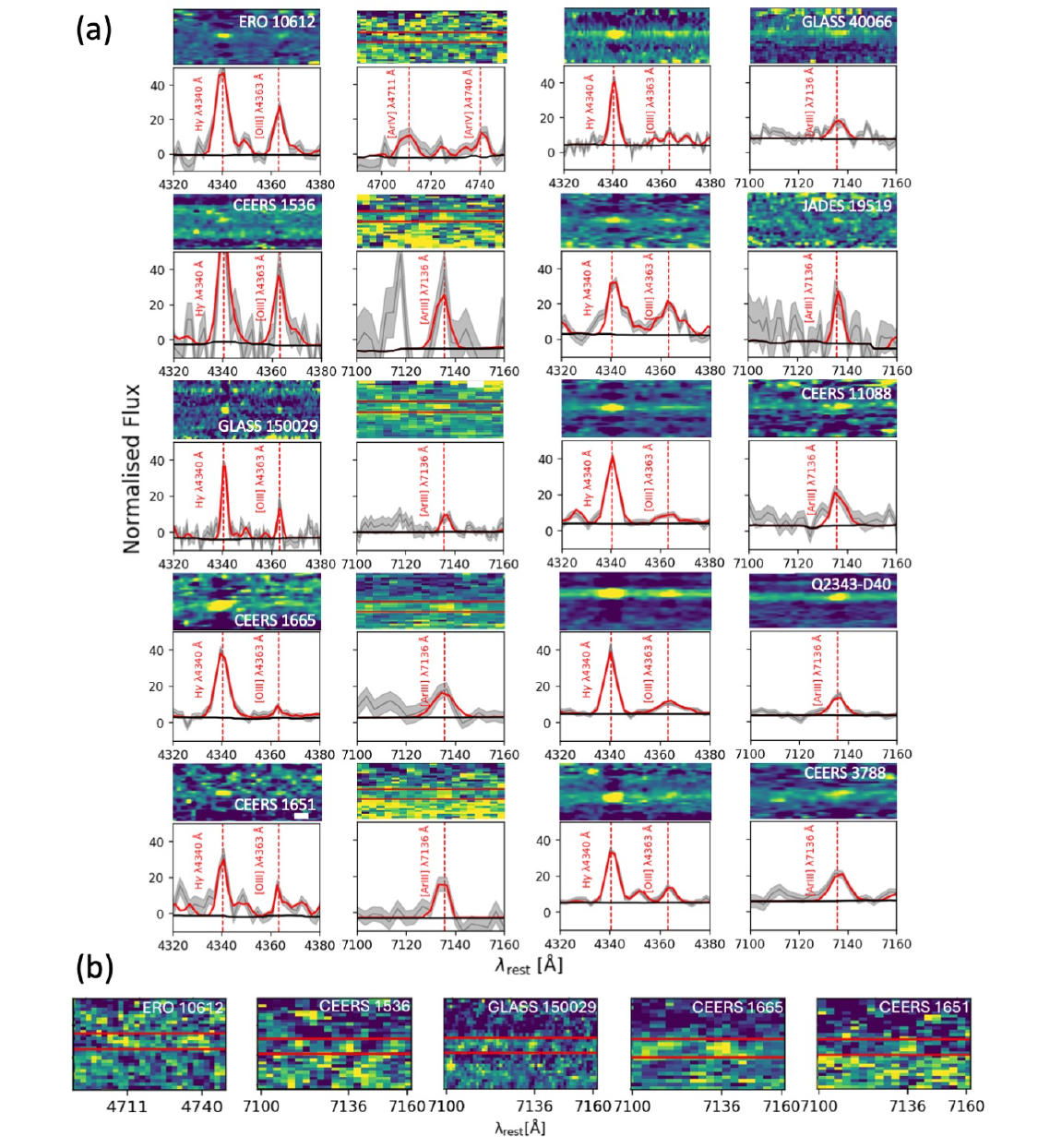}
\caption{(a) [OIII]$\lambda$ 4363 \AA~ and  ionised Ar lines in flux and wavelength calibrated archival 1D JWST/NIRSPEC spectra and in 2D traces, for the ten galaxies analyzed here. Observed 1D spectra are shown in grey (with shaded 1~$\sigma$ uncertainty), best-fit 1D spectra in red, and fitted continuum in black. For the Argon line regions, the color scale of the 2D traces is adjusted to the line fluxes for the highest redshift (see Table~\ref{table:abund}) galaxies (left).  The horizontal red lines enclose the pixels used by the MAST/JWST pipeline for constructing the 1D spectra. (b) The distribution in the 2D spectra of line flux to its error ratio (Flux/$\delta_{Flux}$) around the detected Ar lines for the same five highest-redshift galaxies; color scale is adapted for each 2D trace. Notice in particular that both [ArIV] lines are detected for the $z=7.66$ galaxy ERO 10612 (see also Table~\ref{table:data}) .} 
\label{Fig:lines}
\end{figure*}

\section{Data and abundance determination} 
\label{sec:data1}

\subsection{Emission-line galaxy sample from MAST}
\label{sec:sample1}

We built a sample of high-redshift (z$>$1) galaxies with O and Ar abundances determined directly via the detection of temperature sensitive auroral lines. For this purpose, we first identified galaxies that have O abundances published already in the literature from their [OIII]$\lambda$ 4363 \AA~line detected in JWST/NIRSPEC MOS observations. We then searched for their publicly available\footnote{From the Mikulski Archive for Space Telescopes (MAST) at the Space Telescope Science Institute.} 1D JWST/NIRSPEC grating spectra, and verified, via the publicly available automated line fitting algorithm ALFA software \citep{Wesson16}, whether their [OIII]$\lambda$ 4363 \AA~line was detected with signal-to-noise S/N$>3$ in these spectra, using their previously published redshift as an input parameter. We also checked with ALFA whether the ionised Ar lines (either  [ArIII]$\lambda$ 7136 \AA~ or [ArIV]$\lambda\lambda$ 4711,4740 \AA) were detected with S/N$>3$ in these spectra. For the reliability of the flux and wavelength calibration, please see further in Appendix~\ref{Appendix:calib}.

This procedure resulted in a sample of ten $z>1 $ galaxies for which all the emission lines required for direct abundance determination are present with S/N$>3$, see Table~\ref{table:data}. The selected spectra cover a wide range of exposure times (0.85 – 29.2 hrs) depending on the parent survey; they are shown in Figure~\ref{Fig:lines}a. We note that the spectra of GLASS 150029 and GLASS 40066 have higher spectral resolution than those of the other galaxies, as their observations were carried out with the higher resolution G235H and G395H gratings, while the other galaxies were observed with the medium resolution G235M and G395M gratings. {For the galaxy ERO 10612, two sets of observations were available from the MAST archive, acquired at different telescope rotation angles. For our analysis, we used the exposure with strong emission lines in the spectra that have symmetric line profiles while the other one with asymmetric line profiles is not used.} All the utilized spectra in this study can be accessed via \dataset[DOI:10.17909/yj93-nc36]{https://url.uk.m.mimecastprotect.com/s/683ACDYyMuy6PYXcWflTjmen3?domain=dx.doi.org} from MAST. Two of these galaxies at z$\sim$4.5 (GLASS 150029 and CEERS 1665), having relatively broad H$\alpha$ lines, have been reported as candidate AGN hosts\footnote{As the AGN is unresolved for the two sources, it remains unclear if it is responsible for all their emission-line fluxes. We thus include them in our abundance determination analysis assuming a star-forming ionising source but conservatively they are not considered for the ensuing interpretation.} by \citet{Harikane23} while the other eight are SFGs. 

\begin{figure}[t]
\centering
\includegraphics[width=\columnwidth]{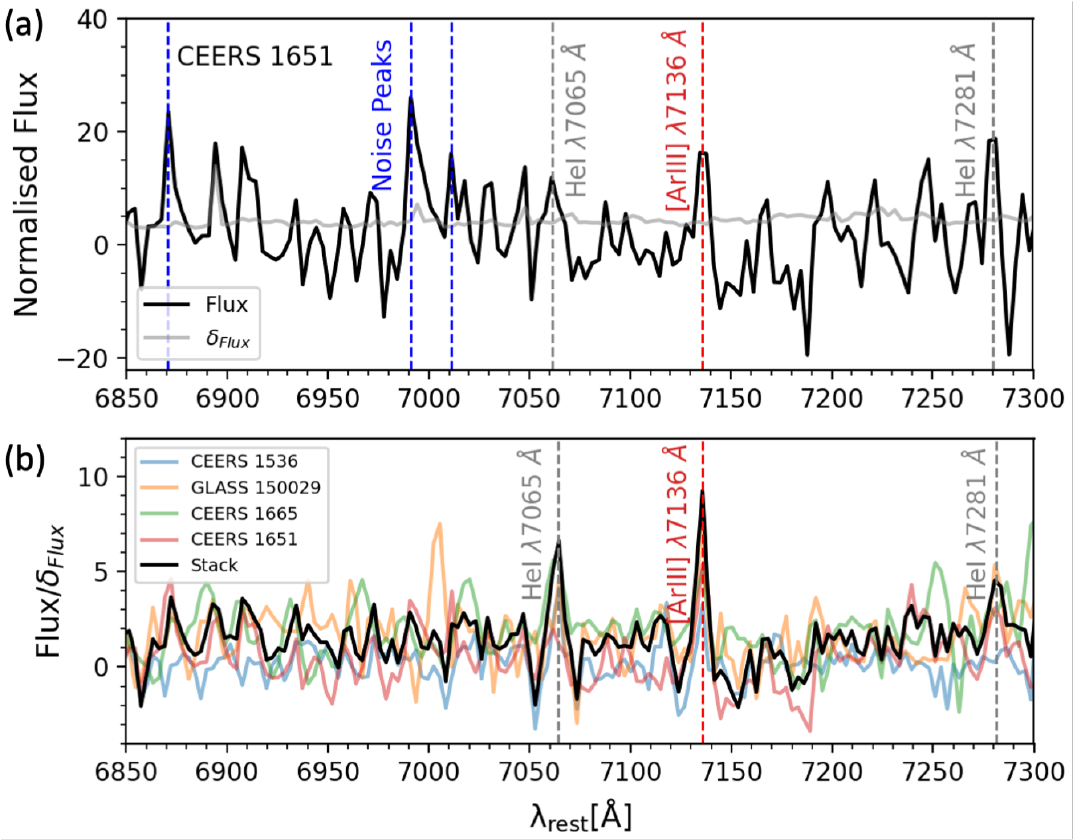}
\caption{(a) 1D spectrum and flux error for CEERS~1651 as a function of rest-frame wavelength, in a broad wavelength range around the detected [ArIII]$\lambda$ 7136 \AA~line. Emission lines with Gaussian like profile such as ionised Ar and He I $\lambda$ 7281 \AA~ line are identified by ALFA, however the He I $\lambda$ 7065 \AA~ is too weak to be identified. Noise peaks, given their non-gaussian profiles, are not identified as emission lines. (b) The Flux/$\delta_{Flux}$ in the 1D spectra plotted as a function of rest-frame wavelength, in a wavelength range around the detected [ArIII]$\lambda$ 7136 \AA~line, for the four highest-redshift galaxies where the line is detected. The Flux/$\delta_{Flux}$ for the 1D stacked spectrum for these four galaxies is marked in black. } 
\label{Fig:stack}
\end{figure}

ALFA measures emission line fluxes from galaxy spectra, after subtracting a globally fitted continuum, by optimizing the parameters of Gaussian fits to the line profiles using a genetic algorithm; line-blending are also taken into account (see Appendix~\ref{Appendix:blend}). For faint lines, the wavelength positions and line profile shape, constrained from stronger emission lines, are used as additional discriminants from noise peaks reaching comparable counts/fluxes. As an example, Figure~\ref{Fig:stack}a shows the 1D spectrum of CEERS 1651 around its [ArIII]$\lambda$ 7136 \AA~line. The ionised Ar line as well as the He I $\lambda$ 7281 \AA~line (while not of interest in this work) have gaussian-like profiles and are detected as lines by ALFA. No line is detected at the position of the He I $\lambda$ 7065 \AA~line (also not of interest in this work) as the S/N at this position is less than 3. Noise peaks showing non-gaussian line profiles are thus rejected by ALFA, even if the S/N at their positions is greater than 3. 

Nevertheless, we will further verify both [OIII]$\lambda$ 4363 \AA~ and ionised Ar line detections by inspecting their 2D spectra as well as via a stacking analysis for the galaxies with faint [ArIII]$\lambda$ 7136 \AA~lines, see Section~\ref{sec:verification1}.

\subsection{Emission line verification}
\label{sec:verification1}

In Figure~\ref{Fig:lines}a, we show the 2D traces of the spectra for the 10 galaxies selected from the MAST archive, at the rest wavelength of the  [OIII]$\lambda$ 4363 \AA~lines and ionised Ar lines. The [OIII]$\lambda$ 4363 \AA~lines are clearly visible in the 2D traces with varying intensity for all galaxies, with the faintest emission seen for CEERS 11088.

The ionised Ar lines are also visible in the 2D traces for all galaxies. We note that for ERO 10612, GLASS 150029 and CEERS 1651 there are noise peaks of similar intensity in the 2D traces. Along with CEERS 1536 \& CEERS 1665, these constitute the highest redshift ($z>3.6$) galaxies in our sample (see Table~\ref{table:abund}). For these five galaxies, we additionally investigate the 2D distribution of their flux to flux error ratio, Flux/$\delta_{Flux}$, around the rest-frame wavelengths of their detected ionised Ar lines, see  Figure~\ref{Fig:lines}b. 

For each galaxy, the Flux/$\delta_{Flux}$ 2D traces display peaks at the wavelengths of the ionised Ar lines ([ArIV] $\lambda\lambda$ 4711,4740 \AA~ for ERO 10612 and [ArIII]$\lambda$ 7136 \AA~ for the others) within the region bracketed by the red lines in Figure~\ref{Fig:lines}b, where the 1D spectrum is extracted. Some off-centered peaks are also visible for CEERS 1536 and CEERS 1665 at $\sim 20-30 $ \AA~ bluer wavelength than the detected Ar line in Figure~\ref{Fig:lines}b. These peaks have sharp profiles which are different from that of the faint line emissions, which are spread on several pixels instead, and better centered within the extraction region.


\begin{deluxetable*}{lcccccccccc}
\tabletypesize{\scriptsize}
\tablewidth{0pt}
\setlength\tabcolsep{3pt}
\tablecaption{Physical parameters and abundances of galaxies with z=1.3--7.7 studied in this work. 
\label{table:abund}}
\tablehead{
\colhead{Name} & \colhead{z} & \colhead{log(M$_{*}$)} & \colhead{log(sSFR)} & \colhead{t$\rm_{SF}$} & \colhead{c(H$\beta$)} & \colhead{T$\rm_{e}$} & \colhead{n$\rm_{e}$} & \colhead{12+log(O/H)} & \colhead{12+log(Ar/H)} & \colhead{log(O/Ar)}
\\
& & [M$_{\odot}$] & [yr$^{-1}$] & [Myr] &  & [K] & [cm$^{-3}$] &  &  & 
}
\startdata 
ERO 10612 & 7.66 	& $7.78 \pm 0.29$ & $-6.64 \pm 0.29$ &	$4.37^{4.13}_{2.13}$ & $0.15 \pm 0.11$	& $18800 \pm 1700$	& $1030 \pm 710$	& $7.73 \pm 0.08$ 	& $5.98 \pm 0.1$ & $1.75 \pm 0.17$\\
CEERS 1536 & 5.038 	& $8.85 \pm 1.09$  & $-7.65 \pm 1.11$ &	$44.67^{530.77}_{41.2}$	& $1.12 \pm 0.19$	& $33600 \pm 5200$	& 1000$^c$	& $7.36 \pm 0.04$ 	& $5.34 \pm 0.14$ & $2.02 \pm 0.15$\\
GLASS 150029$^a$ & 4.584 	& $9.12 \pm 0.33$  & $-8.08 \pm 0.33$ &	$120.23^{136.81}_{63.99}$ & $0.71 \pm 0.11$	& $21100 \pm 2600$	& $286 \pm 285$			& $7.53 \pm 0.08$ 	& $5.51 \pm 0.1$ & $2.02 \pm 0.13$\\
CEERS 1665$^a$ & 4.488 	& $9.79 \pm 0.92$  & $-7.37 \pm 0.92$ 	&	$23.44^{171.54}_{20.62}$ & $1.5 \pm 0.28$	& $12900 \pm 1700$	& $836 \pm 200$	& $8.13 \pm 0.1$ 	& $5.67 \pm 0.11$ & $2.46 \pm 0.15$\\
CEERS 1651 & 4.382 	& $8.85 \pm 0.89$  & $-7.37 \pm 0.9$ 	&	$23.44^{162.77}_{20.49}$ & $1.57 \pm 0.62$	& $20400 \pm 4600$	& 1000$^c$	& $7.49 \pm 0.16$ 	& $5.25 \pm 0.22$ & $2.23 \pm 0.26$\\
GLASS 40066 & 4.02 	& $9.4 \pm 0.31$  & $-7.83 \pm 0.31$ &	$67.61^{70.43}_{34.5}$	& $0.55 \pm 0.04$	& $12700 \pm 800$	& 1000$^c$		& $8.08 \pm 0.06$ 	& $5.65 \pm 0.08$ & $2.4 \pm 0.11$\\
JADES 19519 & 3.604 	& $8.64 \pm 0.1$  & $-8.04 \pm 0.14$ 	&	$109.65^{41.71}_{30.21}$ & $0.94 \pm 0.16$	& $23700 \pm 3400$	& $569 \pm 559$		& $7.48 \pm 0.08$ 	& $5.3 \pm 0.12$ & $2.18 \pm 0.16$\\
CEERS 11088 & 3.302 	& $9.68$  & $-7.35$ &	$22.39$	 & $1.37 \pm 0.08$	& $15700 \pm 2300$	& $606 \pm 152$	& $7.87 \pm 0.11$ 	& $5.45 \pm 0.11$ & $2.42 \pm 0.16$\\
Q2343-D40 & 2.963 	& --  & -- 	& -- & $1.03 \pm 0.1$	& $13000 \pm 1400$	& $68 \pm 67$	& $8.01 \pm 0.09$ 	& $5.5 \pm 0.09$ & $2.5 \pm 0.13$\\
CEERS 3788 & 2.295 	& $9.45$  & $-8.82$ &	$660.69$ & $0.93 \pm 0.3$	& $15300 \pm 1300$	& $197 \pm 53$		& $7.91 \pm 0.07$ 	& $5.56 \pm 0.09$ & $2.35 \pm 0.11$\\
COSMOS 19985$^b$ & 2.188 	& $10.12 \pm 0.04$  & $-7.8 \pm 0.06$ 	&	$63.1^{9.35}_{8.14}$ & $0.52 \pm 0.09$	& $13640 \pm 2900$	& $195 \pm 71$			& $7.89 \pm 0.2^b$ 	& $5.33 \pm 0.24^b$ & $2.58 \pm 0.31^b$\\
COSMOS 20062$^b$ & 2.185 	& $10.1 \pm 0.07$  & $-7.68 \pm 0.08$ 	&	$47.86^{9.68}_{8.05}$ 	& $0.68 \pm 0.13$	& $8900 \pm 2700$	& $231 \pm 64$			& $8.24 \pm 0.27^b$ 	& $5.58 \pm 0.31^b$ & $2.66 \pm 0.41^b$\\
SGAS1723+34$^b$ & 1.329 	& $8.77 \pm 0.15$  & $-7.88 \pm 0.16$ 	&	$75.86^{33.79}_{23.28}$	& $0.07 \pm 0.04$	& $12300 \pm 600$	& $130 \pm 113$			& $8.13 \pm 0.03^b$ 	& $5.69 \pm 0.06^b$ & $2.43 \pm 0.08^b$
\enddata
\tablecomments{Column 1: Name of galaxy; Column 2: galaxy redshift; Column 3-5: Estimated mass, sSFR, and star-formation timescale of galaxies, from their SED fitted to broad-band photometry (see Appendix~\ref{Appendix:mass}); Column 6: Measured Balmer decrement; Column 7: Estimated nebular temperature; Column 8: Estimated electron density; Columns 9--11: Estimated elemental abundances.\\
$^a$Identified from broad H$\alpha$ lines as AGN host by \citet{Harikane23}.\\
$^b$For these galaxies, the abundances were estimated from line fluxes published in the literature. \\
$^c$For these galaxies, the n$\rm_e$ value has been assumed.
}
\end{deluxetable*}

Finally, to further verify the detection of the [ArIII]$\lambda$ 7136 \AA~line fluxes in the publicly available JWST/NIRSPEC spectra for the five highest redshift galaxies in our sample, we computed the stacked 1D spectrum from the individual 1D spectra of CEERS 1536, GLASS 150029, CEERS 1665 and CEERS 1651, adopting the spectral resampling using SpectRes (\citealt{Carnall17}) and normalised to their [ArIII]$\lambda$ 7136 \AA~line fluxes. In the rest frame 1D stacked spectrum, the S/N of a detected line should show an increase proportional to the square root of the number of stacked spectra in case of Poissonian noise \citep[e.g.][]{Arnaboldi02}. The Flux/$\delta_{Flux}$ in the 1D spectra around the [ArIII]$\lambda$ 7136 \AA~lines, for the four galaxies and their 1D spectral stack is shown in Figure~\ref{Fig:stack}b. We find that the [ArIII] $\lambda$ 7136 \AA~line is detected with S/N of 9.27, showing the expected rough increase in S/N compared to the S/N of the line in individual spectra. In the stacked spectrum the He I $\lambda$ 7065 \AA~ line is also clearly seen. 

Note that for ERO 10612 ($z=7.66$), the [ArIII]$\lambda$ 7136 \AA~line lies beyond the red wavelength limit of NIRSPEC, instead both [ArIV]$\lambda\lambda$ 4711, 4740 \AA~ are independently detected, each with S/N$\sim 3-4$, see Table~\ref{table:data}. {In the second exposure for ERO 10612 acquired at a different telescope rotator angle, only the [ArIV]$\lambda$ 4711 \AA~ line is detected with with S/N$\sim3.75$.} The stacked 1D spectra from the two exposures shows that the Flux/$\delta_{Flux}$ of the [ArIV]$\lambda$ 4711 \AA~ line is higher in the stack (S/N$\sim$4.68) than in the individual exposures (3.17 and 3.73) by $\approx \sqrt{2}$, {supporting a robust detection for this line emission}.

\subsection{Flux measurements and final sample}
\label{sec:finalsamp1}

For the 10 galaxies above with extracted 1D JWST/NIRSPEC grating spectra, flux measurements of the detected emission-lines are then carried out using ALFA \citep{Wesson16}. {Therefore noise peaks at random wavelengths are never fit, and the wavelengths of all expected fainter lines are known from the bright lines. All lines are fitted to a Gaussian profile iterated during the fit. Since the emission line profile is approximately Gaussian, while the noise peak profile is generally not, even in the unlikely event that a noise peak falls exactly at the wavelength of an expected emission line, a correspondingly large residual will cause a low final S/N, hence a non-detection.}
The line flux ratios for the emission lines of interest with respect to H$\beta$ are reported in Table~\ref{table:data} for each {of the ten galaxies}, with 1D spectra of {these galaxies} shown in Figure~\ref{Fig:lines}a. 

In addition to these 10 galaxies, we include three high-redshift SFGs in our sample whose relevant line flux measurements, including ionised O and Ar lines, are published in the literature. One SFG at z$\sim$1.3 had its O and Ar abundances determined directly based on their [OIII]$\lambda$ 4363 \AA~line detection in NIRSPEC IFU observations \citep{Welch24}. For two SFGs at z$\sim$2.2, their direct determination of O abundances were based on Keck/MOSFIRE spectra but with temperature sensitive [OII]$\lambda\lambda$ 7322,7332 \AA~line detections,  where [ArIII] $\lambda$ 7136 \AA~line fluxes were also reported \citep{Sanders23}. We use the published line-flux measurements for these three additional SFGs to determine their O \& Ar abundances. 

{In summary, we thus have a sample of 13 $z>1$ emission line galaxies, 3 of them with line flux measurements from the literature, while we measure line fluxes for 10 galaxies from their publicly available 1D JWST/NIRSPEC spectra. For all of them, we determine their  O \& Ar abundances in this work.}

\subsection{Abundance determination} 
\label{sec:abund}

For each galaxy in our sample, the emission-line fluxes, whether taken from published sources or measured from publicly available spectra, are then passed to NEAT \citep[Nebular Empirical Analysis Tool;][]{Wesson12}, which applies an empirical scheme to calculate the extinction and elemental abundances. NEAT calculates the intrinsic c(H$\beta$) using the flux-weighted ratios of H$\alpha$/H$\beta$, H$\gamma$/H$\beta$ and H$\delta$/H$\beta$ (whichever pairs are observed) and the extinction law of \citet{Cardelli89}, first assuming a nebular temperature of 10000K and an electron density of 1000 cm$^{-3}$, and then recalculating c(H$\beta$) at the measured temperature and density.

Emission lines fluxes for each galaxy are de-reddened using the calculated c(H$\beta$), see Table~\ref{table:abund}, and their temperatures and densities are calculated using an iterative process from the relevant diagnostic lines using NEAT \citep[see][section 3.3]{Wesson12}. For our observations, NEAT utilizes the temperature-sensitive [OIII]$\lambda$ 4363 \AA~ line, or the [OII]$\lambda\lambda$ 7322,7332 \AA~lines for the two Keck/MOSFIRE observed galaxies, and the density-sensitive [OII]$\lambda\lambda$ 3726,3729 \AA~, {[ArIV] $\lambda\lambda$ 4711,4740 \AA} and [SII]$\lambda\lambda$ 6717,6731 \AA~ doublets to obtain temperature and electron density for each galaxy spectrum. For three galaxies (see Table~\ref{table:abund}), we do not observe the required doublets to determine electron densities. In these cases, 1000 cm$^{-3}$ is adopted, however the value is expected to have negligible impact on the determined abundances as the emissivities of the auroral line transitions are nearly density independent at the low densities  determined for our observed sources \citep[e.g.][]{Ferland13}. Assuming a lower electron density of 200 cm$^{-3}$, similar to many galaxies in our sample (see Table~\ref{table:abund}) leads to consistently lower log(O/Ar) values of $\sim$0.05 dex and higher 12 + log(Ar/H) values by $\sim$0.03 dex, well within the estimated errors.

O and Ar ionic abundances are measured from the observed fluxes of the O ([OII]$\lambda\lambda$ 3726,3729 \AA~,  [OIII]$\lambda\lambda$ 4363,4959,5007~\AA) and Ar ([ArIII] $\lambda$ 7136 \AA~ and/or [ArIV] $\lambda\lambda$ 4711,4740 \AA) lines respectively. Ionisation correction factor (ICF) for O is negligible when lines pertaining to both O$^{2+}$ (i.e, [OIII] $\lambda\lambda\lambda$ 5007, 4959, 4363 \AA) and  O$^{+}$ ([OII] $\lambda\lambda$ 3727, 3729 \AA) are observed. Elemental Ar abundances are obtained from the Ar$^{2+}$ ionic abundances utilising the ICF from \citet{Amayo21} when [ArIII] $\lambda$ 7136 \AA~ is detected. Only for ERO 10612, the reported Ar abundance is the {Ar$^{3+}$} ionic abundance obtained from [ArIV] $\lambda\lambda$ 4711,4740 \AA~observations with no ICF correction. The impact of the assumed ICF on the determined Ar abundances and thereby the implications for our results are discussed in Appendix~\ref{Appendix:ICF}. Uncertainties are propagated through all steps of the analysis into the final abundance values. Along with the O and Ar abundances in Table~\ref{table:abund}, we report the galaxy stellar masses, specific-star-formation rates (sSFR) and star-formation timescales (t$\rm_{SF}$), which are also are discussed in Appendix~\ref{Appendix:mass}. 


\subsection{Comparison with previously published abundance determinations} 
\label{sec:notes}

\begin{figure}[t]
\centering
\includegraphics[width=\columnwidth]{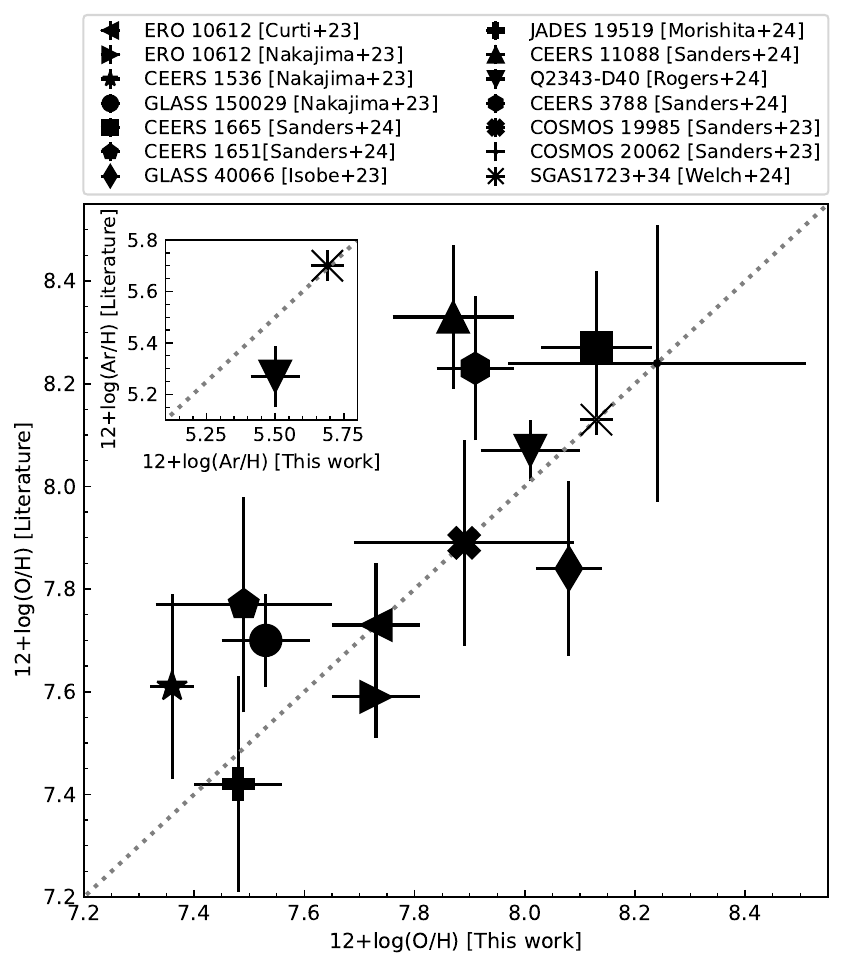}
\caption{12 + log(O/H) abundances determined in this work compared to previously published values. The inset shows the same for the two sources with previously published 12 + log(Ar/H) determinations.} 
\label{Fig:lit}
\end{figure}

In Figure~\ref{Fig:lit} we show our determined O abundances against previously published values. {Most of our galaxies} (\citealt{Curti23}: ERO 10612; \citealt{Sanders24}: CEERS 1665, \citealt{Isobe23}: GLASS 40066; \citealt{Morishita24}: JADES 19519; \citealt{Rogers24}: Q2343-D40; \citealt{Sanders23}: COSMOS 19985, COSMOS 20062; \citealt{Welch24}: SGAS1723+34)  are consistent within the errors, but some offset is observed for a few others (\citealt{Nakajima23}: ERO 10612, CEERS 1536, GLASS 150029; \citealt{Sanders24}: CEERS 1651, CEERS 11088, CEERS 3788). Our newly determined 12 + log(O/H) values have a mean offset of $0.09$~dex and standard deviation of $0.18$~dex compared to the literature values. The small differences are mainly {related} to differences in flux and wavelength calibration between our utilised archival spectra and those utilised by previous studies. This is illustrated for ERO 10612 where our determined O abundance is consistent with that reported by \citet{Curti23} but not with that reported by \citet{Nakajima23}, with both these authors using different in-house spectral flux and wavelength calibrations. Indeed, applying our flux measurement and abundance determination procedure to older versions of JWST NIRSPEC spectra for our studied galaxies, as released by previous authors\footnote{Particularly public versions of the spectra for \href{https://ceers.github.io/dr07.html}{CEERS DR0.7} and \href{https://jades-survey.github.io/scientists/data.html}{JADES DR3} were analysed.}, we find our determined 12 + log(O/H) values have a lower mean offset of $0.05$~dex and lower standard deviation of $0.06$ dex compared to literature works.  

We report in this work the first determination of Ar abundances and thereby log(O/Ar) for all these galaxies except SGAS1723+34 and Q2343-D40. For Q2343-D40, \citet{Rogers24} reported an Ar abundance slightly lower than our determination, a potential consequence of their choice for a different ICF scheme \citep{Izotov06} than that utilised here, or arising from differences in their abundance determination methodology. For SGAS1723+34, \citet{Welch24} report an Ar abundance nearly identical to our determination (see Figure~\ref{Fig:lit} [inset]). 


\begin{figure*}[ht!]
\centering
\includegraphics[width=0.98\columnwidth]{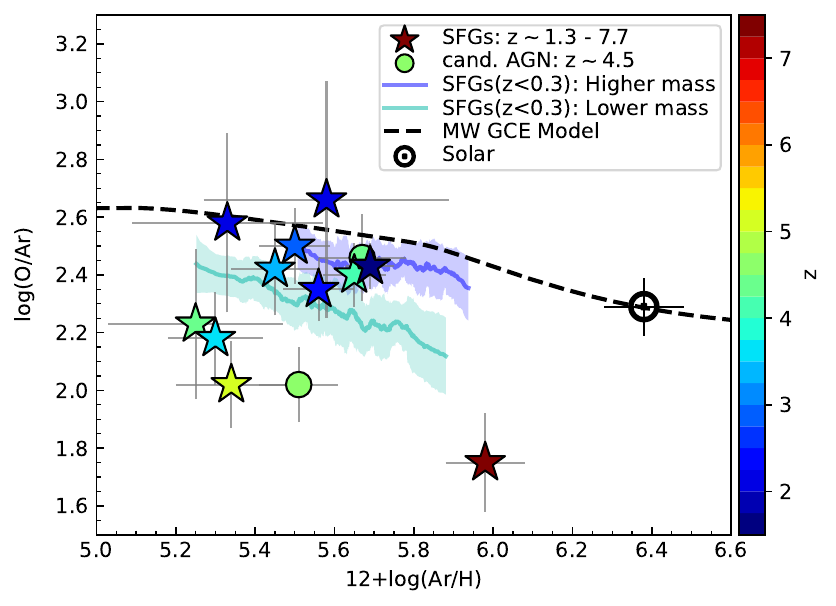}
\includegraphics[width=\columnwidth]{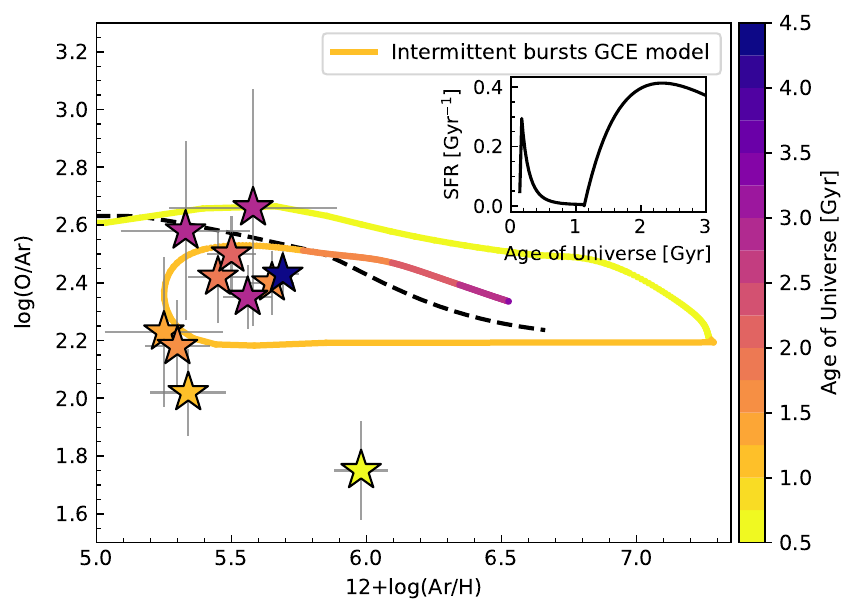}
\caption{[Left] log(O/Ar) vs 12+log(Ar/H) for the 11 SFGs at z$\sim$1.3--7.7 and two candidate AGN hosts at z$\sim$4.5. The galaxies are coloured by their redshift. The green and blue lines respectively show the sequence of mean values of low-redshift (z$<0.3$) relatively lower mass ({$<$log(M$\rm_{*}$/M$\rm_{\odot}$)$>$ = 7.23}) and higher mass ({$<$log(M$\rm_{*}$/M$\rm_{\odot}$)$>$ = 9.41}) galaxies from SDSS (Bhattacharya et al. in prep). Their standard deviations are shaded. The MW solar neighbourhood GCE model \citep{kob20sr} is shown as black dashed line. [Right] Same as [Left] but now the SFGs are coloured by the age of the universe at their redshift and the AGN candidates are not plotted. Also plotted is a GCE model for an intermittent starburst scenario, that follows the star-formation history shown in the inset, coloured by age of the universe.} 
\label{Fig:jwst}
\end{figure*}

\section{Galaxy chemical enrichment at z$\sim$1.3--7.7} 
\label{sec:enrich}

Figure~\ref{Fig:jwst} [Left] shows the position of the galaxies at z$\sim$1.3--7.7 in the log(O/Ar) vs 12 + log(Ar/H) plane, and represents their state of chemical enrichment. 

Seven SFGs (z$\sim$1.3--4; see Figure~\ref{Fig:jwst} [Left]) are consistent within error with the Milky Way (MW) solar neighbourhood Galactic Chemical Evolution (GCE) model \citep{kob20sr}, where CCSNe (including hypernovae) and SNe Ia dominate in a self-regulated scenario with no inflows or outflows. Additionally, their position is consistent with the locus traced by the mean log(O/Ar) as a function of 12+log(Ar/H) for the higher mass ({$<$log(M$\rm_{*}$/M$\rm_{\odot}$)$>$ = 9.41}) low-redshift (z$<0.3$) starbursts from SDSS \citep[][Bhattacharya et al. in prep]{Brinchmann04}. This remarkably shows that, just like the MW ISM \citep{Matteucci21,kob20sr} and higher-mass starbursts at z$<0.3$, the SFGs in our sample out to z$\sim$4 are consistent with having chemical enrichment being driven mainly by CCSNe and SNe Ia, with the same nucleosynthesis yields and initial mass functions. 

{We note that O \& Ar abundance determination from JWST/NIRSPEC spectra have now been reported independently for an additional 8 galaxies at z$\sim$1.8--5.2 by \citet{Stanton24} and 4 more at z$\sim$3.2--4.7 by \citet{Stiavelli24}. Almost all these galaxies have log(O/Ar) and 12+log(Ar/H) values consistent with the MW GCE model (see also Figure 5 in \citealt{Stanton24}).} 

The only other SFG (KBSS-LM1, z=2.396, \citealt{Steidel16}) with direct auroral-line measurements and full-spectral UV-fitting\footnote{Other SFGs with [Fe/H] determined from full-spectral UV-fitting \citep[e.g.][]{Cullen19, Cullen21, Stanton24} have [O/H] (in lieu of $\alpha$ abundance) determined from strong-line methods from optical emission-lines, which may be over-estimated if not calibrated against direct methods \citep{Maiolino19}.} has [O/Fe]$\sim 0.6 \pm 0.13$ at [Fe/H]$\sim$-1.6. This is coincident with the MW solar neighbourhood GCE model in the [O/Fe] vs [Fe/H] plane (see Fig~3 in \citealt{kob20sr}), and in extension consistent with the {z$\sim$1.3--4 SFGs} presented in this work. 

On the other hand, four SFGs (z$\sim$3.5--7.7) have log(O/Ar) values below the MW GCE model considering also their errors and scatter, and even below the sequence traced by lower-mass (M$<10^8$ M$_{\odot}$) z$<0.3$ starbursts (Figure~\ref{Fig:jwst} [Left]). O and Ar have similar dust condensation temperatures \citep{Savage96}, therefore it is unlikely that their log(O/Ar) value is underestimated on account of preferential ejection of only condensed O with dust grains.

Therefore in our small sample of galaxies MW-like self-regulated chemical enrichment sequences and their underlying mechanisms hold up to z$\sim$4. At z$>$3.5 or so, SFGs appear to be characterised by higher Ar abundance relative to O. In the following subsections, we discuss possible scenarios that may be responsible for the low log(O/Ar) values for the higher redshift galaxies in our sample.


\subsection{Potential chemical enrichment from intermittent starbursts at high redshift} 
\label{sec:model}

Keeping the same assumption of the chemical enrichment recipe for the solar neighbourhood \citep{kob20sr}, we construct {an illustrative} GCE model considering two bursts of star-formation at very early times (with substantial infall of pristine gas between the bursts), to try and explain the position of the four low log(O/Ar) SFGs in this plane (see Figure~\ref{Fig:jwst} [Right]). An intermittent star-formation model has previously been invoked to explain measurements of emission line fluxes in similar high redshift SFGs \citep{Kobayashi24}. Note that the purpose of constructing this specific model is to {investigate whether} the low log(O/Ar) values of these galaxies may be explained by considering an extreme star-formation history {while} keeping the MW-like CCSNe and SNe Ia dominated nucleosynthesis. 

In Figure~\ref{Fig:jwst} [Right] (inset), the star-formation history of the GCE model is plotted. An initial burst of star formation at the break of cosmic dawn is followed by a quiescent phase up to $\sim$1.1 Gyr after the birth of the universe, when the ISM is enriched through continued explosions of Type Ia SNe to very high 12 + log(Ar/H) values and minimum log(O/Ar) values. This is followed by an infall of primordial gas that strongly dilutes the ISM, reducing 12 + log(Ar/H) keeping log(O/Ar) constant, and induces another episode of star-formation that starts re-enriching the ISM. Thus a loop in the log(O/Ar) vs 12 + log(Ar/H) plane (Figure~\ref{Fig:jwst} [Right]) is formed which is a signature of gas infall \citep{Spitoni19,Arnaboldi22,Kobayashi23}. Other models with less extreme star-formation histories (lower star formation rates with continuous or intermittent star-formation with and without gas infall) will occupy the parameter space within the loop spanned by this extreme model in this plane. The model can successfully explain the positions of three of the SFGs at z$\sim$3.5--5 that have log(O/Ar)$\sim2.1$. Note that SFGs are not expected to appear on the constant log(O/Ar)$\sim$2.1 line, except for at very low 12+log(Ar/H) values, for the model (Figure~\ref{Fig:jwst} [Right]) as this signifies the rapid dilution of the ISM and star-formation only follows afterwards. 

We note however that ERO 10612, which is the highest redshift galaxy in our sample at z=7.66, i.e, just 672 Myr after the birth of the universe\footnote{We assume Planck cosmological parameters \citep{Planck20}: Hubble constant H$_0$ = $67.4 \pm 0.5$ km s$^{-1}$ Mpc$^{-1}$, matter density parameter $\Omega\rm_{m}$ = $0.315 \pm 0.007$. With these assumptions we determine the age of the universe at given redshifts for the galaxies in our sample in Figure~\ref{Fig:jwst} [Right]. We assume epoch of first star formation at z=25, within the expected z=20--30 range \citep{Bromm09}. This has been applied to the GCE model shown in Figure~\ref{Fig:jwst} [Right] (inset).}, has a log(O/Ar) value {$\sim$1.8}, that is below the log(O/Ar) values reached by the intermittent star-formation model (see Figure~\ref{Fig:jwst}). Additional physical mechanisms, such as outflows, may be at play for this galaxy. A possible scenario may be {one where after} the first burst of star-formation the CCSNe ejecta with O and Ar {are} expelled in an outflow while Ar produced over longer timescales is retained, eventually reaching the observed log(O/Ar) {value $\sim$1.8}. 

It remains to be demonstrated whether CCSNe and SNe Ia dominated chemical enrichment models could explain the low-log(O/Ar) value for ERO 10612, especially given the short timespan after the birth of the universe within which such abundance values need to be reached. {Future} GCE simulations for this galaxy exploring different star-formation histories with inflows and outflows may address this property. {Additional sources of chemical enrichment described in literature sources may also potentially be at play, as discussed concisely in the following section.} 

\subsection{Additional potential sources of chemical enrichment at high redshift} 
\label{sec:potential}

A potential mechanism for additional Ar production leading to log(O/Ar)-poor stellar populations would be to assume a higher SN Ia rate (e.g. with a higher binary fraction) relative to CCSNe for a given stellar population mass in the early universe, than is seen for the MW solar neighbourhood. 

Another potential mechanism for Ar enhancement would be the inclusion of sub-Chandrasekhar mass SNe Ia (sub-Ch SNe Ia; see \citealt{kob20ia} and references therein) in the GCE models. These sub-Ch SNe Ia have been suggested to be the main enrichment source for observed MW dwarf spheroidal satellite galaxies, which formed their masses at early times and have been quiescent since \citep{Kirby19}. GCE models including sub-Ch SNe Ia \citep{kob20ia} predict very low log(O/Ar), as well as low [$\alpha$/Fe], because of the higher occurrence of sub-Ch SNe Ia relative to Ch-mass SNe Ia at early times (after $\sim$40~Myr). 

Another potential rapid metal enrichment source is pair-instability supernovae \citep[PISNe;][]{Heger02,Nomoto13} that have been predicted to evolve from massive ($>140$~M$_{\odot}$) population III stars (first generation metal-free stars formed from pristine gas) having only $\sim2$~Myr \citep{Takahashi18} lifespans. However, their metal contribution to the ISM is expected to be visible only for a short time after the first generation of stars formed \citep{Hartwig18,Vanni23}, to be washed-out rapidly \citep{Ji15} once CCSNe enrichment processes begin after $\sim20$~Myr \citep{kob20sr}. Given the small, albeit uncertain, t$\rm_{SF}$ value of ERO 10612 (see Table~\ref{table:abund}), PISNe could in principle explain its low log(O/Ar) values based on model PISN O and Ar yields \citep{Takahashi18}, but only if there has been no mixing with any pre-enriched ISM or mass-loss from pre-existing stars. This possibility underlines the interest in further detailed studies of ERO 10612 and {other} potentially similar systems. However, in GCE models with conventional assumptions, the PISN enrichment causes the rapid decrease of log(O/Ar) to very low values at much lower metallicities (Kobayashi et al. in prep). 


\section{Summary and conclusions} 
\label{sec:conc}

We extend the use of the log(O/Ar) vs 12 + log(Ar/H) plane \citep{Arnaboldi22,Kobayashi23} for inferring the mechanisms that govern galaxy chemical enrichment to SFGs, offering a direct analogy to the [$\alpha$/Fe] vs [Fe/H] plane for stars. We robustly obtain line-flux measurements for {eight} SFGs from their flux and wavelength calibrated 1D JWST/NIRSPEC spectra (see Table~\ref{table:data}), in addition to {three} SFGs where such flux measurements were available from their literature sources (see Section~\ref{sec:data1}). We then directly determine O and Ar abundances for these 11 SFGs at z$\sim$1.3--7.7 from observations of temperature sensitive auroral lines (Table~\ref{table:abund}). 

We present their positions in the log(O/Ar) vs 12 + log(Ar/H) plane (Figure~\ref{Fig:jwst}a). {Seven} SFGs (z$\sim$1.3--4) are consistent within error with a MW-like CCSNe and SNe Ia dominated chemical enrichment model \citep{kob20sr}. {Four} SFGs (z$\sim$3.5--7.7) are found to be log(O/Ar)-poor compared {to the other seven aforementioned SFGs and} the MW GCE model track (Figure~\ref{Fig:jwst}).

Thus, in the majority of our small sample of galaxies, MW-like self-regulated chemical enrichment sequences and their underlying mechanisms may be in place as early as z$\sim$1.3--4. {This is corroborated for independent galaxy samples in \citet{Stanton24} and \citet{Stiavelli24}.} 

The low log(O/Ar) values of three SFGs at higher redshift (z$\sim$3.5--5.) may be explained through a tailored GCE model with early intermittent star-formation, but keeping the MW-like CCSNe and SNe Ia dominated chemical enrichment (see Section~\ref{sec:model}). For ERO 10612 (z=7.66), {deeper observations would improve the S/N of the faint [ArIV]$\lambda$ 4740 \AA~line. Exploration} of different star-formation histories, potentially with more bursts, considering also inflows and outflows that preferentially eject O and/or additional potential sources of Ar enrichment may be required {to explain the low log(O/Ar) determined for these galaxies}. Dedicated GCE models will be utilized to better understand the rapid Ar enrichment of these SFGs (Kobayashi et al. in prep). 

The ever-improving quality of JWST/NIRSPEC data and upcoming large ground-based spectroscopic surveys \citep[e.g. the Prime Focus Spectrograph galaxy evolution survey at Subaru;][]{pfs}, implies that it should be possible to build-up a large sample of SFGs with direct determinations of O and Ar abundances from auroral line-flux measurements. In conjunction with tailored GCE models, such a large sample of galaxies with elemental abundance determinations will enable further refinement in the understanding of galaxy chemical enrichment mechanisms presented in this work. There is, thus, now a new window for constraining galaxy chemical enrichment from present day to the early universe. 
 
\begin{acknowledgments}
We thank the anonymous referee for their very useful comments. SB was supported by the INSPIRE Faculty award (DST/INSPIRE/04/2020/002224), Department of Science and Technology (DST), Government of India.
SB and MAR acknowledge support to this research from the European Southern Observatory, Garching, through the 2022 SSDF. SB and OG acknowledge support to this research from Excellence Cluster ORIGINS, which is funded by the Deutsche Forschungsgemenschaft (DGF, German Research Foundation) under Germany's Excellence Strategy - EXC-2094-390783311. MAR and OG thank the Research School of Astronomy and Astrophysics at ANU for support through their Distinguished Visitor Program in 2024. This work was supported by the DAAD under the Australia-Germany joint research programme with funds from the Australian Ministry for Science and Education. CK acknowledges funding from the UK Science and Technology Facility Council through grant ST/Y001443/1.
\end{acknowledgments}

\vspace{5mm}
\facilities{JWST (NIRSPEC), Keck (MOSFIRE), SDSS}

\software{{AstroPy \citep{astropy:2013, astropy:2018, astropy:2022}}, SciPy \citep{scipy}, NumPy \citep{numpy}, Matplotlib \citep{matplotlib}, SpectRes \citep{Carnall17}, NEAT \citep{Wesson12} and ALFA \citep{Wesson16}.}


\appendix

\section{Reliability of flux and wavelength calibrations for JWST/NIRSPEC spectra}
\label{Appendix:calib}

The spectra are flux calibrated with an approximately 15\% absolute flux accuracy, while field-dependent variations may be as large as 10\% as per the \href{https://jwst-docs.stsci.edu/jwst-calibration-status/nirspec-calibration-status/nirspec-mos-calibration-status}{latest JWST/NIRSPEC MOS calibration pipeline}. Emission line fluxes estimated from medium resolution grating spectra seem to consistently have $\sim$10\% higher flux values compared to PRISM spectra and NIRCAM photometry \citep{Bunker23}, although line flux ratios (as used here) seem unaffected. As per the same latest calibrations, wavelength calibration is accurate till $\sim$15 km/s and $\sim$40 km/s for high and medium resolution gratings respectively. Detector artifacts have been more effectively removed by the pipeline for the latest available spectra.


\section{Impact of line-blending on Argon abundances}
\label{Appendix:blend}

The majority of the O and Ar lines used in this work are not expected to be blended with other lines. Only the [ArIV]$\lambda$ 4711 \AA~line, seen only for ERO~10612 in our sample, may occasionally be blended with the HeI $\lambda$ 4713 \AA~line. The genetic fitting algorithm of ALFA fits the observed spectra simultaneously to a number of lines, including the [ArIV]$\lambda$ 4711 \AA~ and HeI $\lambda$ 4713 \AA~lines. Given the spectral resolution and the observed flux distribution $\sim$4711 \AA~ for ERO~10612, any observed flux was determined by ALFA to be attributed solely to [ArIV]$\lambda$ 4711 \AA~, with no contribution from the HeI $\lambda$ 4713 \AA~line (if both lines had contributed to the observed flux, then the flux distribution would have been broader). We note that even if these two lines are blended, and the determined Ar abundance of ERO 10612 is overestimated, any potential correction will result in reducing the Ar abundance while increasing its log(O/Ar) by the same value, thereby still keeping the same diagonal offset in Figure~\ref{Fig:jwst} from the MW-like chemical enrichment sequence in the log(O/Ar) vs 12 + log(Ar/H) plane.


\section{Impact of ionisation correction factors on Argon abundances}
\label{Appendix:ICF}

For Ar, the ionisation states of Ar$^{2+}$, Ar$^{3+}$, and in smaller amounts Ar$^{+}$ and Ar$^{4+}$ are possible for an SFG. Hence the ICF correction to the observed ionic abundances is relevant for Ar abundance determination.  \citet{Cordova24} found that different ICF schemes, including \citet{Amayo21} used here, have worked equally well for Ar abundance determination. They found that for z$\sim$0.1 SFGs with 12 + log(O/H)$>$8.2, ICF correction from observing only [ArIII] $\lambda$ 7136 \AA~may underestimate the Ar abundance by up to $\sim0.4$~dex in their sample, but no such effect is seen when 12 + log(O/H)$<$8.2. This implies that the resulting log(O/Ar) values would be higher than their actual intrinsic values. The ICF correction based on the observation of both [ArIII] and [ArIV] lines is more accurate instead at all 12 + log(O/H) values (see their Figure 3). Only one of our SFGs (COSMOS~20062) has 12 + log(O/H)$>$8.2 with its Ar abundance value determined from the observed Ar$^{2+}$ lines only. Its Ar abundance value may indeed be underestimated but even considering the maximum offset, its log(O/Ar) abundance would still be consistent with MW-like chemical enrichment, as it is the galaxy with the highest log(O/Ar)  in our sample (see Table~\ref{table:abund}). Other SFGs {with [ArIII] $\lambda$ 7136 \AA~ detection} in our sample have 12 + log(O/H)$<$8.2 and thus no ICF correction bias is expected. 

{For ERO 10612, we have not utilised any ICF correction as no literature ICF scheme directly provides a recipe for Ar abundance determination when only  Ar$^{3+}$ ionic abundance is determined. Even if such a scheme were to come to pass, the Ar abundance of ERO 10612 would only increase by a given value, leading to a reduction in log(O/Ar) by the same value, thereby still keeping the same diagonal offset in Figure~\ref{Fig:jwst} from the MW-like chemical enrichment sequence in the log(O/Ar) vs 12 + log(Ar/H) plane.} 


\section{Stellar mass, specific-star-formation-rate and star-formation-timescales of the galaxy sample}
\label{Appendix:mass}

For all the galaxies with z$>4$, the stellar mass and sSFR (noted in Table~\ref{table:abund}) were reported by \citet{Nakajima23}, based on their JWST Spectral Energy Distribution (SED) fitting with assumed initial mass function (IMF) from \citet{Chabrier03}. For JADES 19519, the stellar mass and sSFR based on JWST SEDs was reported by \citet{Morishita24}. For CEERS 11088 and 3788, we instead report the Hubble Space Telescope SED- based \citep{Momcheva16} estimates of stellar mass and sSFR with the same assumed IMF. The stellar mass and sSFR for COSMOS 19985 and 20062 \citep{Sanders23} and SGAS1723+34 \citep{Florian21} are also noted in Table~\ref{table:abund} but these estimates are not available for Q2343-D40. Star-formation timescale is computed with the simple approximation of t$\rm_{SF}=1/sSFR$. We note that a top-heavy IMF would reduce the stellar masses of the galaxies by $\sim$0.5~dex \citep{Harvey24}.

\bibliography{oar}{}
\bibliographystyle{aasjournal}

\end{document}